\documentclass{article}
\usepackage{graphicx} 
\usepackage{amsmath}
\usepackage{amssymb}
\usepackage{hyperref}
\hypersetup{colorlinks,linkcolor={blue},citecolor={blue},urlcolor={blue}}
\usepackage[version=4]{mhchem}
\usepackage{mathtools, nccmath}
\usepackage{setspace}
\usepackage{etoc}
\usepackage{cite}
\usepackage{float}
\usepackage{lipsum}
\usepackage{blindtext}
\usepackage[table,xcdraw]{xcolor}
\usepackage{multirow}
\usepackage{booktabs}
\usepackage{siunitx}
\usepackage{booktabs}
\usepackage{outline}
\usepackage{avant} 
\usepackage{multirow}
\usepackage[table,xcdraw]{xcolor}

\usepackage{array}
\usepackage{rotating}
\usepackage{epstopdf}
\usepackage{tikz}
\usepackage{lscape}
\usepackage{subcaption}
\usepackage[toc,page]{appendix}
\usepackage{enumitem}

\title{The eV-Scale Sterile Neutrino and Neutrinoless Double Beta Decay}

\author{{Priya}{\footnote{kashyappriya963@gmail.com}, Simran Arora\footnote{009simranarora@gmail.com}, and B. C. Chauhan\footnote{chauhan@associates.iucaa.in}}}

\date{\textit{Department of Physics and Astronomical Science, Central University of Himachal Pradesh, Dharamshala (HP) 176215, India}}

\begin{document}
    \maketitle  
    \begin{abstract}

       In short-baseline experiments such as LSND and MiniBooNE, an excess of electron neutrinos has been observed, originating from a muon neutrino beam. To address this anomaly, in the line of many works, we investigate various neutrino mixing schemes involving eV-scale sterile neutrinos alongside three active neutrinos. Using updated experimental and global fit data, we studied neutrinoless double beta decay for three different schemes such as 3+1, 1 + 3, and 2 + 2, which involve one sterile neutrino and three active neutrinos. We have done analysis of these schemes for normal hierarchy (NH) as well as for inverted hierarchy (IH) frameworks, and constrained the sterile neutrino mass in light of current and future neutrinoless double beta decay experiments. The 3+1 scheme is found to be the most viable and at the level of $3\sigma$ the mass of sterile neutrino with respect to the lightest neutrino mass ($m_{\text{lightest}}$) is restricted to $4.75~eV$ for the NH and $4.72~eV$ for the IH. Additionally, the limits on the sum of four neutrino masses are determined to be $4.81~eV$ for the normal hierarchy and $4.78~eV$ for the inverted hierarchy. The updated analysis of all these schemes would help us in understanding physics governing neutrinoless double beta decay and limit on the mass of sterile neutrinos.
    \end{abstract}


\section{Introduction}
The purpose of neutrinoless double beta decay  (0$\nu\beta\beta$) studies is far more basic than just measuring neutrino mass; it is the search for lepton number violation and for Majorana nature of the neutrinos \cite{1,2}. A special case of double beta decay known as `neutrinoless double beta decay' occurs when no neutrinos (or antineutrinos) are produced as products\cite{3}. The reaction governing 0$\nu\beta\beta$ is given by
\begin{gather*}
(A, Z) \rightarrow (A, Z+2) + 2e^-,
\end{gather*}
where the lepton number is violated by 2 units $(\Delta L = 2)$\cite{5}. A number of experiments are now underway or are scheduled to begin soon in the search of neutrinoless double beta decay\cite{6,7,KamLAND-Zen:2024eml,8,9,10,11,12,13,14}. The KamLAND-Zen experiment currently holds the best position in establishing the experimental limit for 0$\nu\beta\beta$, which search the decay in $ ^{136}Xe$-doped liquid scintillator. It has established a lower limit for the $0\nu\beta\beta$ decay half life of $\tau \geq 3.8$ × $10^{26}$ years at a 90\% confidence level (CL). This provides upper limit of effective Majorana mass parameter ($m_{\beta\beta}$) in the range $28-122$ meV at 90\% CL\cite{14,KamLAND-Zen:2024eml}.

In the Standard Model of particle physics, there are three known types of neutrinos: the electron neutrino, muon neutrino, and tau neutrino. These are called 'active' neutrinos because they interact with other particles through the weak force and oscillate due to the mass squared differences, for example, of the order $10^{-5} eV^2$ and $10^{-3} eV^2$ \cite{15}. Many different scientific evidences have been pointing towards the possible existence of something called `sterile neutrinos' at a specific energy scale \cite{16}. These hints come from various experiments and observations, including the Liquid Scintillator Neutrino Detector (LSND) \cite{17,18}, Gallium experiments \cite{19,20}, T2K experiment \cite{21}, and the MiniBooNE experiment \cite{22,23,24}. The LSND experiment was a short-baseline experiment with the baseline length $\sim$ 30m, that investigates $\bar{\nu}_\mu \to \bar{\nu}_e$  oscillations and reported an excess of $\bar{\nu}_e$-like events\cite{Gninenko:2010pr} . It observed an event excess with a significance of 3.8$\sigma$. Although, a similar experiment, KARMEN, with the baseline length $\sim$ 17m also studied the oscillation channel $\bar{\nu}_\mu \to \bar{\nu}_e$, but did not confirm the LSND anomaly\cite{Eitel:2000by,Drexlin:2003fc}.

The MiniBooNE experiment, designed to confirm the LSND anomaly, also reports a low-energy excess in the electron like events. The MicroBooNE experiment was established to test this excess using a liquid argon time projection chamber \cite{MicroBooNE:2016pwy}. The first three years of MicroBooNE data show no evidence of sterile neutrino osillations and are consistent with the 3$\nu$ hypothesis at the $1\sigma$ level \cite{MicroBooNE:2022sdp,MicroBooNE:2025khi}. However, see Ref. \cite{Denton:2021czb}, when focused on the disappearance of electron neutrinos across four channels in MicroBooNE data, a hint of oscillations was observed. A strong signal (2.4$\sigma$) was observed in the Wire-Cell analysis, with preferred parameters $\sin^2(2\theta_{14}) = 0.35^{+0.19}_{-0.16}$ and $\Delta m^2_{41} = 1.25^{+0.74}_{-0.39} \, \text{eV}^2$.

The Baksan Experiment on Sterile Transitions (BEST) investigated the gallium anomaly and reported best fit parameters \( \Delta m^2 = 3.3^{+\infty}_{-2.3} \,\text{eV}^2 \) and \( \sin^2 2\theta = 0.42^{+0.15}_{-0.17} \), consistent with SAGE and GALLEX with higher significance \cite{Barinov:2022wfh}. Similarly, GALLEX/GNO and SAGE observed this anomaly at a \(3.2\sigma\) confidence level \cite{Serebrov:2022ajm}. However, the NEUTRINO-4 experiment reported evidence for active-sterile oscillations at the \( 2.9\sigma \) level with best-fit parameters \( \Delta m^2_{14} = (7.3 \pm 1.17) \,\text{eV}^2 \) and \( \sin^2 2\theta = 0.36 \pm 0.12 \) \cite{Serebrov:2020kmd}.  On the other hand, several high-precision reactor antineutrino experiments contradict these claims \cite{Naumov:2021vds}. The DANSS experiment found no statistically significant evidence for sterile neutrinos, with the best-fit point in the 4\(\nu\) case having only \(1.3\sigma\) significance \cite{Danilov:2021oop}. The NEOS experiment also reported no strong evidence for 3+1 neutrino oscillations, constraining the mixing parameter to \( \sin^2 2\theta_{14} < 0.1 \) for \( \Delta m^2_{41} \) in the range 0.2–2.3 \(\text{eV}^2\) at a 90\% confidence level \cite{NEOS:2016wee}. Similarly, the Double Chooz experiment found no indication of sterile neutrino mixing and set exclusion limits in the mass range $5 \times 10^{-3}~\text{eV}^2 \lesssim \Delta m^2_{41} \lesssim 3 \times 10^{-1}~\text{eV}^2$, for mixing angle $\sin^2 2\theta_{14} \gtrsim 0.02$, further constraining the allowed parameter space for light sterile neutrinos \cite{DoubleChooz:2020pnv}. The RENO experiment provides the most stringent limits on sterile neutrino mixing in the region $\Delta m^2_{41} \lesssim 0.002~\text{eV}^2$, based on precise measurements of electron antineutrino ($\bar{\nu}_e$) disappearance \cite{RENO:2020uip}. The Daya Bay experiment has excluded light sterile neutrino mixing with $\sin^2 2\theta_{14} \gtrsim 0.01$ at the 95\% confidence level in the mass range $0.01~\text{eV}^2 \lesssim \Delta m^2_{41} \lesssim 0.1~\text{eV}^2$ \cite{DayaBay:2024nip}

The investigation of sterile neutrinos may shed light on why ordinary neutrinos have mass. Three scales of neutrino mass squared differences $\Delta m^2_{\text{solar}}$, $\Delta m^2_{\text{atm}}$ \cite{25} and $\Delta m^2_{\text{LSND}}$\cite{26} have been found through solar, atmospheric neutrino oscillation experiments and LSND collaboration. These three mass-squared differences are required to explain the active-sterile neutrino framework. This type of work has been widely explored in the literature \cite{Goswami:2024ahm,Goswami,27,28,29,30,31,32,33,34,35,36,37,Dekens,Barry,Bilenky}.

In our work, motivated by the updated global data of short baseline neutrino experiments, reactor experiments, and updates of the experiments on the nature and mass scale of neutrinos, we have investigated the third mass squared difference in the 3 $\sigma$ range, i.e., $1.1-2.4$ $eV^2$ and at the best-fit value, i.e., 1.73 $eV^2$ \cite{27}. We also constrained the mass of light sterile neutrinos and the sum of four neutrino masses in light of data from the current and future generations of experiments.

 The most general case involving sterile neutrinos is discussed in the literature, where three active neutrinos are considered as light mass eigenstates and the sterile neutrino is regarded as a heavier mass eigenstate. The lightest neutrino mass, effective Majorana mass, mass of sterile neutrino and sum of neutrino masses have also been constrained in various works. Readers can refer to \cite{29,30,31} for detailed study.

In this work, we have considered different active-sterile schemes for analysis of effective Majorana mass in 0$\nu \beta \beta$ in light of current and future experiments. We have taken three possible schemes with three active neutrinos and one sterile neutrino. We have identified the most viable scheme by exploring the third mass squared difference and obtained constraints on the mass of sterile neutrinos. Based on our current understanding of neutrino mass squared differences and mixing angles it is possible to analyse the range of values for the effective Majorana mass \(m_{\beta\beta}\) in neutrinoless double-beta decay as a function of the absolute scale of neutrino masses\cite{26}. 

The paper is organised as follow: In Section 2, we discuss about three different active-sterile neutrino schemes and checked the viability in light of existing bounds. In Section 3, we discuss the effective Majorana mass in 3+1 scheme for normal and inverted hierarchies. The details of the numerical analysis and interpretation of the phenomenological results are discussed in Section 4. Finally, in Section 5 the conclusions are elucidated.

\maketitle

\section{Different Possible Active-Sterile Neutrino Schemes}

\begin{figure}[H]
     \centering
     \begin{subfigure}[b]{0.3\textwidth}
         \centering
         \includegraphics[width=\textwidth]{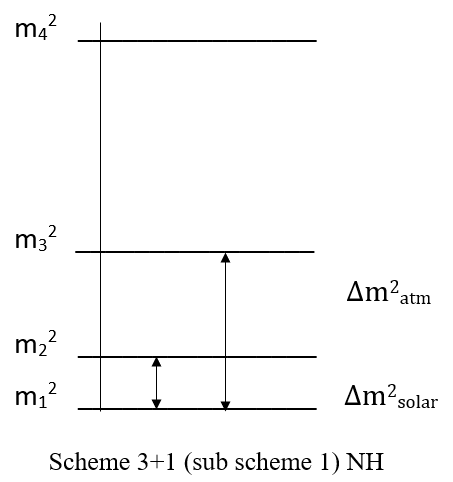}
         \caption{}
        
     \end{subfigure}
     \hfill
     \begin{subfigure}[b]{0.3\textwidth}
         \centering
         \includegraphics[width=\textwidth]{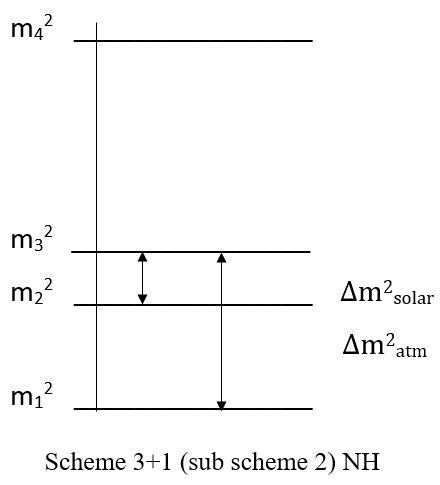}
         \caption{}
         
     \end{subfigure}
     \hfill
     \begin{subfigure}[b]{0.3\textwidth}
         \centering
         \includegraphics[width=\textwidth]{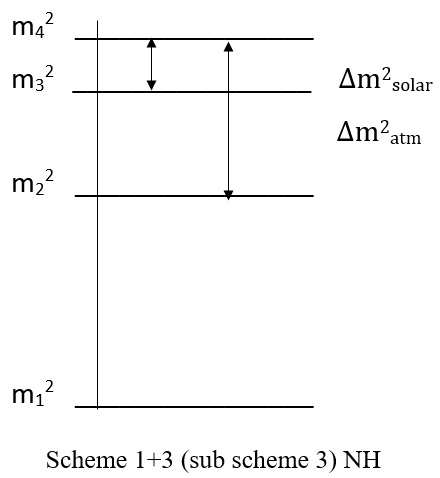}
         \caption{}
         
     \end{subfigure}
      \hfill  

     \begin{subfigure}[b]{0.3\textwidth}
         \centering
         \includegraphics[width=\textwidth]{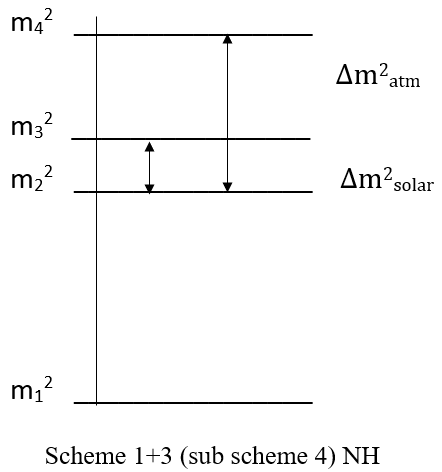}
         \caption{}
         
     \end{subfigure}
     \hfill
     \begin{subfigure}[b]{0.3\textwidth}
         \centering
         \includegraphics[width=\textwidth]{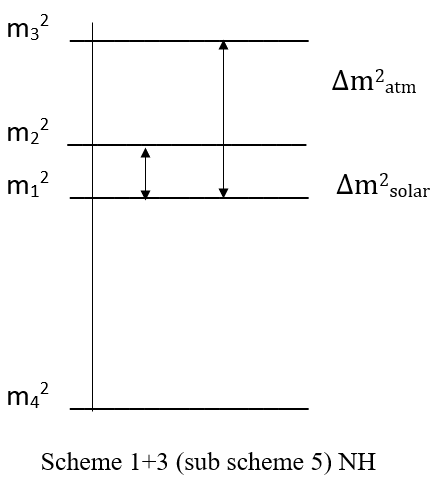}
         \caption{}
         
     \end{subfigure}
     \hfill
     \begin{subfigure}[b]{0.3\textwidth}
         \centering
         \includegraphics[width=\textwidth]{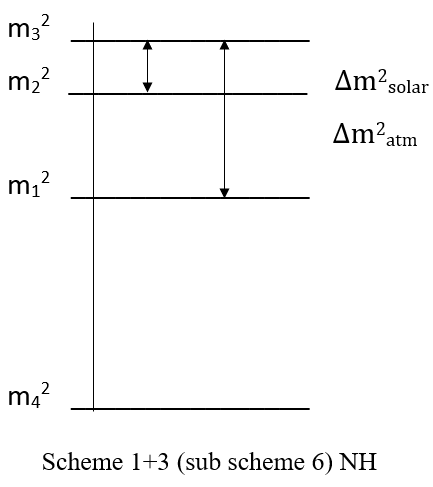}
         \caption{}
         
     \end{subfigure}
     \hfill
     \begin{subfigure}[b]{0.3\textwidth}
         \centering
         \includegraphics[width=\textwidth]{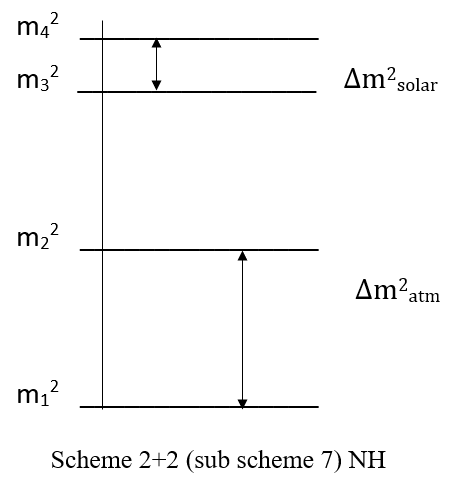}
         \caption{}
         
     \end{subfigure}
     \hfill
     \begin{subfigure}[b]{0.3\textwidth}
         \centering
         \includegraphics[width=\textwidth]{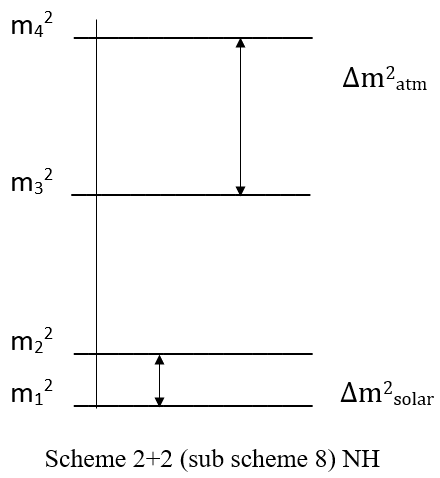}
         \caption{}
         
     \end{subfigure}
        \caption{Different Sub schemes in 3+1, 1+3 and 2+2 Schemes (NH)}
\label{s1}       
\end{figure}

In this section, we describe different neutrino mixing schemes possible through the various combination of four neutrinos i.e. three active neutrinos and one sterile neutrino: 3+1, 1+3, and 2+2 \cite{32}. In the 3+1 scheme, three active neutrinos behave as light mass eigenstates, while the sterile neutrino acts as a heavy mass state. We have explored possible mass squared differences between the active neutrinos in sub schemes 1 and 2. For the 1+3 scheme, one light mass eigenstate is accompanied by three heavy mass eigenstates. Sub schemes 3 and 4 consider the sterile neutrino as a heavy mass eigenstate along with two active neutrinos, while sub schemes 5 and 6 involve the sterile neutrino as an extra light mass eigenstate, with the three active neutrinos being heavier than this light mass eigenstate shown in Fig. \ref{s1}. In 2+2 scheme, two active neutrino behave as light mass and one active and one sterile neutrino behave as heavy mass eigenstates (sub schemes 7 and 8).

\begin{figure}[H]
     \centering
     \begin{subfigure}[b]{0.3\textwidth}
         \centering
         \includegraphics[width=\textwidth]{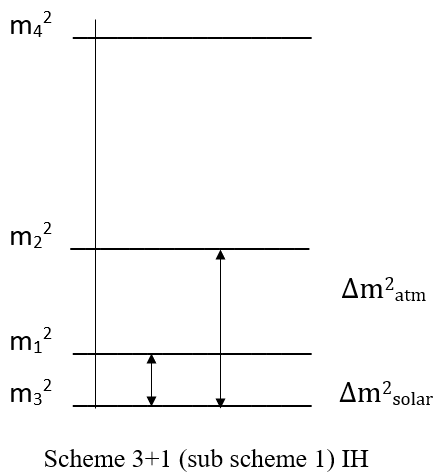}
         \caption{}
         
     \end{subfigure}
     \hfill
     \begin{subfigure}[b]{0.3\textwidth}
         \centering
         \includegraphics[width=\textwidth]{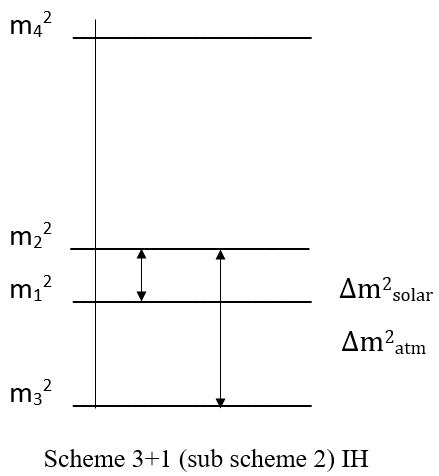}
         \caption{}
         
     \end{subfigure}
     \hfill
     \begin{subfigure}[b]{0.3\textwidth}
         \centering
         \includegraphics[width=\textwidth]{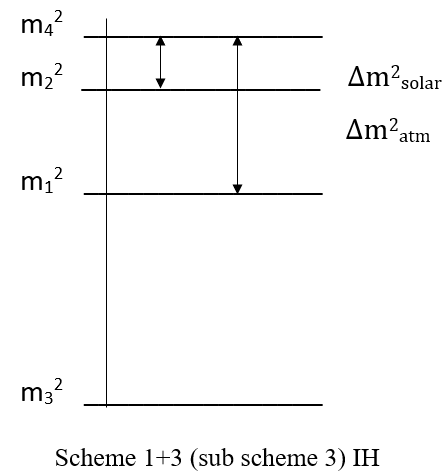}
         \caption{}
         
     \end{subfigure}
     \hfill
     \centering
     \begin{subfigure}[b]{0.3\textwidth}
         \centering
         \includegraphics[width=\textwidth]{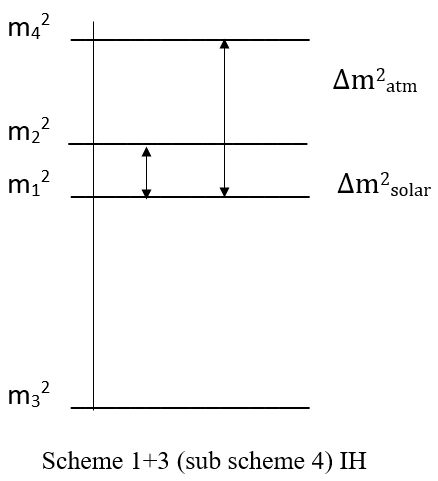}
         \caption{}
         
     \end{subfigure}
     \hfill
     \begin{subfigure}[b]{0.3\textwidth}
         \centering
         \includegraphics[width=\textwidth]{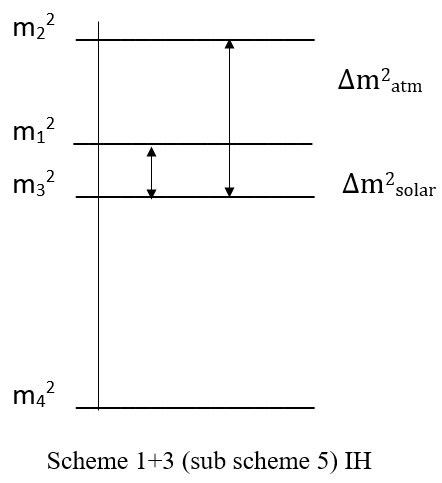}
         \caption{}
         
     \end{subfigure}
     \hfill
     \begin{subfigure}[b]{0.3\textwidth}
         \centering
         \includegraphics[width=\textwidth]{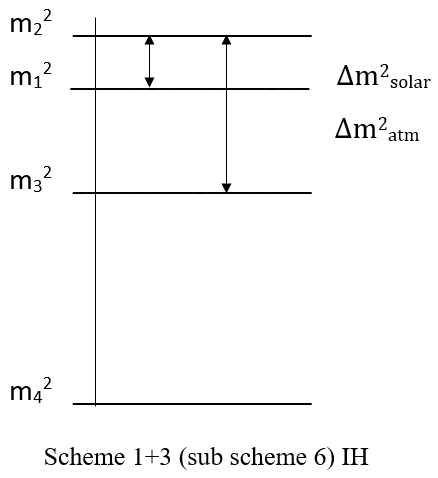}
         \caption{}
         
     \end{subfigure}
     \hfill
     \begin{subfigure}[b]{0.3\textwidth}
         \centering
         \includegraphics[width=\textwidth]{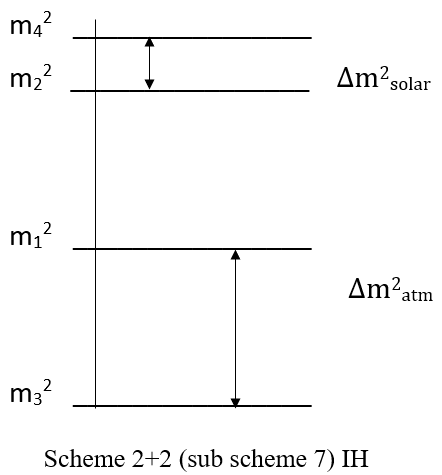}
         \caption{}
         
     \end{subfigure}
     \hfill
     \begin{subfigure}[b]{0.3\textwidth}
         \centering
         \includegraphics[width=\textwidth]{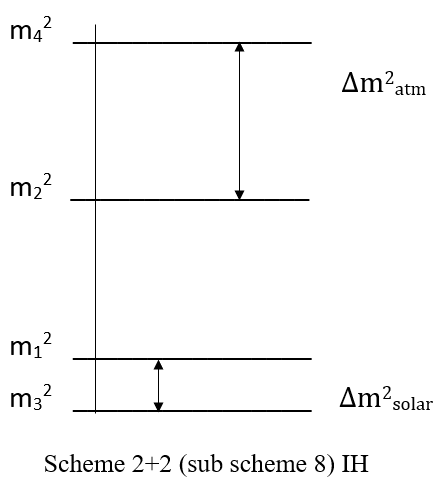}
         \caption{}
         
     \end{subfigure}
        \caption{Different Sub schemes in 3+1, 1+3, 2+2 Schemes (IH)}
        
\end{figure}

In 1+3 scheme (3, 4), the sterile neutrino oscillates with one active neutrino and mass squared differences are of the order of $10^{-5} eV^2$ and $10^{-3} eV^2$, respectively. However, the LSND anomaly suggests that the third mass squared difference is approximately of the order of 1 $eV^2$. If we consider the oscillations of one active and sterile neutrino with mass squared differences of order of $10^{-5} eV^2$ and $10^{-3} eV^2$, respectively then the sum of masses of three active neutrinos is disallowed by cosmological bounds\cite{39}. Similarly, the 2+2 scheme (7, 8) is also incompatible for the same reason \cite{33}. The sub schemes 5 and 6 of 1+3 scheme are not allowed by the LSND anomaly because the extra light mass state sterile neutrino and the active neutrino cannot have a mass squared difference of the order of 1 $eV^2$.

Finally for the most viable 3+1 scheme, the mass squared differences $\Delta m^2_{\text{solar}}$ and $\Delta m^2_{\text{atm}}$ are shown below in the tabular form with NH and IH for sub scheme 1 and 2, as

\begin{table}[h]
    \centering
    \renewcommand{\arraystretch}{1.5}
    \begin{tabular}{|c|c|}
        \hline
        \textbf{Sub scheme} & \textbf{Mass Squared Differences} \\
        \hline
        NH-1 & $\Delta m^2_{21} = \Delta m^2_{\text{solar}},\quad \Delta m^2_{31} = \Delta m^2_{\text{atm}}$ \\
        NH-2 & $\Delta m^2_{32} = \Delta m^2_{\text{solar}},\quad \Delta m^2_{31} = \Delta m^2_{\text{atm}}$ \\
        IH-1 & $\Delta m^2_{13} = \Delta m^2_{\text{solar}},\quad \Delta m^2_{23} = \Delta m^2_{\text{atm}}$ \\
        IH-2 & $\Delta m^2_{21} = \Delta m^2_{\text{solar}},\quad \Delta m^2_{23} = \Delta m^2_{\text{atm}}$ \\
        \hline
    \end{tabular}
    \caption{Mass-squared differences in the 3+1 neutrino sub schemes}
    \label{tab:3plus1}
\end{table}

\section{Effective Majorana Mass in 3+1 Schemes}

The effective Majorana mass for three neutrinos is defined by the following equation

\begin{equation}
    m_{\beta\beta}^{3\nu} = \sum_{j=1}^{3} U^2_{e j} m_j = \sum_{j = 1}^3 |U_{ej}|^2 e^{2 i \phi_j}m_j,
\label{eqn1}
\end{equation}

where $ U_{e j}$ are the elements of PMNS matrix and $\phi_j = \text{arg}~ U_{ej}$. The inclusion of the sterile neutrino introduces modifications to the mixing angle and alterations in the Pontecorvo-Maki-Nakagawa-Sakata (PMNS) matrix as

\begin{equation}
    U = R_{34}\tilde R_{24}\tilde R_{14}R_{23}\tilde R_{13}R_{12}P,
\end{equation}

where the matrices $R_{ij}$ are rotations in the $ij$ space, e.g.

\[\tilde R_{14} = \begin{pmatrix}
    c_{14} & 0 & 0 & s_{14}e^{-i\delta_{14}} \\
    0 & 1 & 0 & 0 \\
    0 & 0 & 1 & 0 \\
    -s_{14}e^{i\delta_{14}} & 0 & 0 & c_{14} \\
\end{pmatrix},
\]

\[R_{34} = \begin{pmatrix}
    1 & 0 & 0 & 0 \\
    0 & 1 & 0 & 0 \\
    0 & 0 & c_{34} & s_{34} \\
    0 & 0 & -s_{34} & c_{34}\\
\end{pmatrix}.
\]

Here $\delta{ij}$, $s_{ij}$ and $c_{ij}$ are the Dirac CP- violating phases, $\sin\theta_{ij}$ and $\cos\theta_{ij}$, respectively. The matrix $P$ is described by the three Majorana phases $\alpha$, $\beta$, and $\gamma$ as 

\[P =  \begin{pmatrix}
    1 & 0 & 0 & 0 \\
    0 & e^{i\alpha/2} & 0 & 0 \\
    0 & 0 & e^{i(\beta/2 + \delta_{13)}} & 0 \\
    0 & 0 & 0 & e^{i(\gamma/2+\delta_{14})} \\
\end{pmatrix}.
\]

There are total three Dirac CP-voilating phases $\delta_{ij}$. The P matrix constructed in such a way that only Majorana phases will show up in the effective Majorana mass governing neutrinoless double beta decay \cite{Barry}. By extending the Eq. (\ref{eqn1}), the effective Majorana mass formula for four neutrinos is

\begin{equation}
m_{\beta\beta}^{4\nu} = \sum_{j=1}^{4} U^2_{e j} m_j,
\end{equation}

\begin{equation}
m_{\beta\beta}^{4\nu} = |U_{e1}|^2 m_1 + |U_{e2}|^2 m_2 e^{i\alpha} + |U_{e3}|^2 m_3 e^{i\beta} + |U_{e4}|^2 m_4 e^{i\gamma}.
\label{eqn3}
\end{equation}
In Eq. (\ref{eqn3}) $m_1,m_2,m_3,m_4$ are the masses of neutrinos. After using the values of $U_{e1}, U_{e2}, U_{e3},$ and $U_{e4}$ from PMNS matrix, the Eq. (\ref{eqn3}) is modified as 

\begin{equation}\label{meff}
\begin{split}
    m_{\beta\beta}^{4\nu} = & \left[ (c_{13}^2 c_{12}^2 c_{14}^2 m_1 + c_{13}^2 s_{12}^2 c_{14}^2 m_2 \cos\alpha + s_{13}^2 c_{14}^2 m_3 \cos\beta + s_{14}^2 m_4 \cos\gamma)^2 \right. \\
    & + \left. (c_{13}^2 s_{12}^2 c_{14}^2 m_2 \sin\alpha + s_{13}^2 c_{14}^2 m_3 \sin\beta + s_{14}^2 m_4 \sin\gamma)^2 \right]^{1/2}.
\end{split}
\end{equation}
For normal hierarchy, $m_1$ is the lightest mass and for inverted hierarchy, $m_3$ is the lightest mass. The forms of effective Majorana mass for four neutrino schemes in NH and IH are as follows

\subsection{3+1 Scheme (NH)}
For sub scheme 1

\begin{equation}
\begin{split}
    m_{\beta\beta}^{4\nu} = & \left[ (c_{13}^2 c_{12}^2 c_{14}^2 m_1 + c_{13}^2 s_{12}^2 c_{14}^2 \sqrt{\Delta m_{solar}^2 + m_1^2} \cos\alpha + s_{13}^2 c_{14}^2 \sqrt{\Delta m_{atm}^2 + m_1^2} \cos\beta \right. \\
    & + \left. s_{14}^2 \sqrt{\Delta m_{41}^2 + m_1^2} \cos\gamma)^2 + (c_{13}^2 s_{12}^2 c_{14}^2 \sqrt{\Delta m_{solar}^2 + m_1^2} \sin\alpha
    \right. \\
    & + \left. s_{13}^2 c_{14}^2  \sqrt{\Delta m_{atm}^2 + m_1^2} \sin\beta + s_{14}^2 \sqrt{\Delta m_{41}^2 +m_1^2} \sin\gamma)^2 \right]^{1/2}.
\end{split}
\end{equation}

For sub scheme 2

\begin{equation}
\begin{split}
    m_{\beta\beta}^{4\nu} = & \left[ (c_{13}^2 c_{12}^2 c_{14}^2 m_1 + c_{13}^2 s_{12}^2 c_{14}^2 \sqrt{\Delta m_{atm}^2  - \Delta m_{solar}^2 + m_1^2} \cos\alpha + s_{13}^2 c_{14}^2 \sqrt{\Delta m_{atm}^2 + m_1^2} \cos\beta \right. \\
    & + \left. s_{14}^2 \sqrt{\Delta m_{41}^2 +m_1^2} \cos\gamma)^2 + (c_{13}^2 s_{12}^2 c_{14}^2 \sqrt{\Delta m_{atm}^2  - \Delta m_{solar}^2 + m_1^2} \sin\alpha
    \right. \\
    & + \left. s_{13}^2 c_{14}^2  \sqrt{\Delta m_{atm}^2 + m_1^2} \sin\beta + s_{14}^2 \sqrt{\Delta m_{41}^2 +m_1^2} \sin\gamma)^2 \right]^{1/2}.
\end{split}
\end{equation}

The forms of effective Majorana mass for different schemes in IH are as follows

\subsection{3+1 Scheme (IH)}
For sub scheme 1

\begin{equation}
\begin{split}
    m_{\beta\beta}^{4\nu} = & \left[ (c_{13}^2 c_{12}^2 c_{14}^2 \sqrt{\Delta m_{solar}^2 + m_3^2} + c_{13}^2 s_{12}^2 c_{14}^2 \sqrt{\Delta m_{atm}^2 + m_3^2} \cos\alpha + s_{13}^2 c_{14}^2 m_3 \cos\beta \right. \\
    & + \left. s_{14}^2 \sqrt{\Delta m_{43}^2 + m_3^2} \cos\gamma)^2 + (c_{13}^2 s_{12}^2 c_{14}^2 \sqrt{\Delta m_{atm}^2 + m_3^2} \sin\alpha
    \right. \\
    & + \left. s_{13}^2 c_{14}^2 m_3 \sin\beta + s_{14}^2 \sqrt{\Delta m_{43}^2 + m_3^2} \sin\gamma)^2 \right]^{1/2}.
\end{split}
\end{equation}

For sub scheme 2

\begin{equation}
\begin{split}
    m_{\beta\beta}^{4\nu} = & \left[ (c_{13}^2 c_{12}^2 c_{14}^2 \sqrt{m_{atm}^2 - \Delta m_{solar}^2+m_3^2} + c_{13}^2 s_{12}^2 c_{14}^2 \sqrt{\Delta m_{atm}^2 + m_3^2} \cos\alpha + s_{13}^2 c_{14}^2 m_3 \cos\beta \right. \\
    & + \left. s_{14}^2 \sqrt{\Delta m_{43}^2 +m_3^2} \cos\gamma)^2 + (c_{13}^2 s_{12}^2 c_{14}^2 \sqrt{\Delta m_{atm}^2 + m_3^2} \sin\alpha
    \right. \\
    & + \left. s_{13}^2 c_{14}^2  m_3 \sin\beta + s_{14}^2 \sqrt{\Delta m_{43}^2 +m_3^2} \sin\gamma)^2 \right]^{1/2}.
\end{split}
\end{equation}

\section{Results and Discussions}

The $m_{\beta\beta}$ as calculated by using Eq. (\ref{meff}) clearly depends on mass eigenvalues, neutrino mixing angles and Majorana phases $\alpha$, $\beta$ and $\gamma$. We used the values of these parameters as given in Table \ref{tab:neutrino_data}. In the computation, the lightest neutrino mass $m_{\text{lightest}}$ i.e., $m_1$ for normal hierarchy and $m_{3}$ for inverted hierarchy are allowed to vary randomly in the range ($10^{-5} - 1$) $eV$. The mass of sterile neutrino ($m_{4}$) is calculated corresponding to the $m_{\text{lightest}}$. The unknown Majorana phases ($ \alpha, \beta$ and $\gamma$) are freely allowed to vary randomly from \(0^\circ\)- \(360^\circ\). The third mass squared difference is explored randomly in the range ($1.1-2.4$) $eV^2$ and at the best fit value, i.e., 1.73\cite{27}. The third mass-squared difference, \( \Delta m^2_{14} = (7.3 \pm 1.17) \,\text{eV}^2 \), reported by the NEUTRINO-4 experiment, is also considered in this study.

We have used the $1 \sigma$ level, $3 \sigma$ level and best fit values for the parameters $\Delta m_{\text{solar}}^2$, $\Delta m_{\text{atm}}^2$, $\theta_{12}$, $\theta_{13}$ and $\theta_{14}$ as per given in the Table 2.

\begin{table}[H]
\footnotesize
    \centering
    \renewcommand{\arraystretch}{1} 
    \begin{tabular}{l c c c c} 
        \hline
        Parameter & best-fit$\pm1\sigma$ range (NH) & best-fit$\pm1\sigma$ range (IH) & $3\sigma$ range (NH) & $3\sigma$ range (IH) \\
        \hline
        $\sin^2\theta_{12}$ & $0.307^{+0.012}_{-0.011}$ & $0.308^{+0.012}_{-0.011}$ & $0.275 - 0.345$ & $0.275 - 0.345$ \\
        $\sin^2\theta_{23}$ & $0.561^{+0.012}_{-0.015}$ & $0.562^{+0.012}_{-0.015}$ & $0.430 - 0.596$ & $0.437 - 0.597$ \\
        $\sin^2\theta_{13}$ & $0.02195^{+0.00054}_{-0.00058}$ & $0.02224^{+0.00056}_{-0.00057}$ & $0.02023 - 0.02376$ & $0.02053 - 0.02397$ \\
        $\Delta m^2_{atm}/ 10^{-3} \text{eV}^2$ & $2.534^{+0.025}_{-0.023}$ & - $2.510^{+0.024}_{-0.025}$ & $2.463 - 2.606$ & $-2.584 - -2.438$ \\
        $\Delta m^2_{solar}/ 10^{-5} \text{eV}^2$ & $7.49^{+0.19}_{-0.19}$ & $7.49^{+0.19}_{-0.19}$ & $6.92 - 8.05$ & $6.92 - 8.05$ \\
        $\Delta m_{4s}^2 $/$eV^2$ & 1.73 &  1.73 &  1.1 - 2.4 & 1.1 - 2.4  \\
        \hline
    \end{tabular}
    \caption{The neutrino oscillation data from global fits used in the numerical analysis. The parameter s in $\Delta m_{4s}^2$/$eV^2$ takes the value s = 1 for NH and s= 3 for IH \cite{Esteban:2024eli,27}}
    \label{tab:neutrino_data}
\end{table}

\begin{figure}[H]

\begin{subfigure}{0.5\textwidth}
  
  \includegraphics[width=1\linewidth]{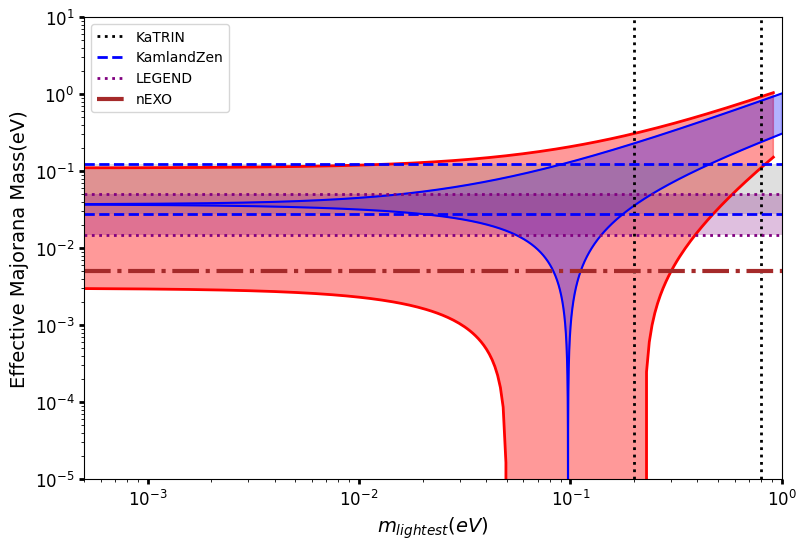}
  \caption{}
\end{subfigure}%
\hfill
\begin{subfigure}{0.45\textwidth}
  
  \includegraphics[width=1\linewidth]{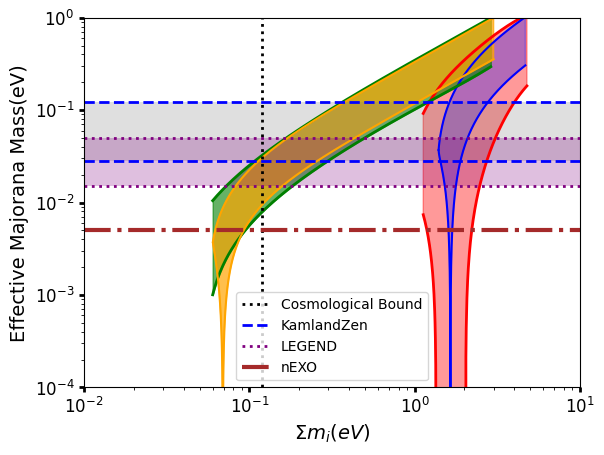}
  \caption{}
\end{subfigure}%

\begin{subfigure}{0.5\textwidth}
  
  \includegraphics[width=1\linewidth]{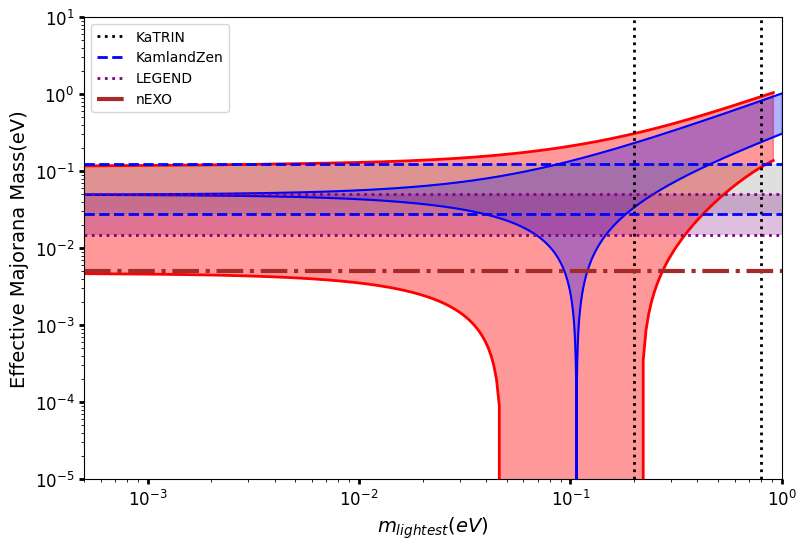}
  \caption{}
\end{subfigure}%
\hfill
\begin{subfigure}{0.45\textwidth}
  
  \includegraphics[width=1\linewidth]{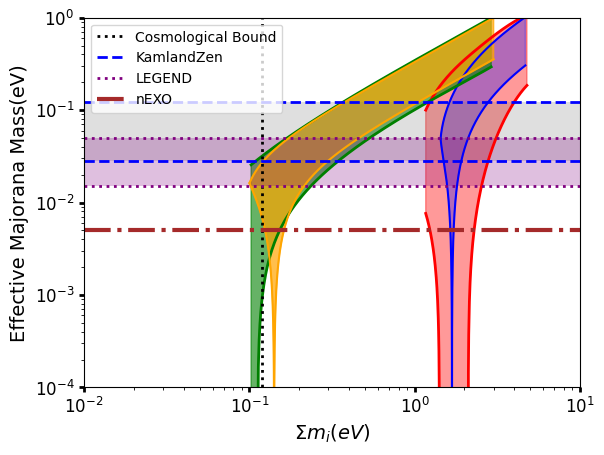}
  \caption{}
\end{subfigure}

\caption{Effective Majorana mass as a function of lightest mass $m_{\text{lightest}}$ (a, c) and sum of masses $\Sigma$ (b, d) for 3+1 scheme (NH) for 1 $\sigma$ and 3 $\sigma$ ranges}
\label{fig3}
\end{figure}

The variation of $m_{\beta\beta}$ as a function of $m_{\text{lightest}}$ and sum of neutrino masses $\Sigma$ in the 3+1 scheme for normal hierarchy is shown in Fig. (\ref{fig3}). The horizontal lines denote the experimental ranges of different experiments like KamLAND-Zen and GERDA\cite{6,7,8}, CUORE \cite{9,10}, LEGEND\cite{11} and nEXO \cite{12}.

The vertical lines in the figures (left panels) denote the current and future ranges for $m_{\text{lightest}}$, given by the KATRIN experiment i.e. 0.8 eV (current limit) and 0.2 eV (future limit) \cite{40}. These lines are included in plots illustrating the $m_{\beta\beta}$ as a function of $m_{\text{lightest}}$. However, the vertical line in the figures (right panels) depicts the cosmological bounds on the sum of three neutrino masses\cite{39}. These lines are included in plots illustrating the $m_{\beta\beta}$ as a function of sum of neutrino masses i.e. $\Sigma$.

The analysis of Fig. (\ref{fig3}) (a) shows the effective Majorana mass, i.e., $m_{\beta\beta}$ for sub scheme 1 (NH) is consistent with present and future experimental data ranges. It can be observed from Fig. (\ref{fig3}) (b) that the sum of four neutrino masses at 3$\sigma$ level is limited to the interval between ($1.11 - 4.81$) eV (depicted by red region) and the sum of four neutrino masses at 1$\sigma$ level is limited to the interval between ($1.41 - 4.71$) eV (depicted by blue region). The sum of three neutrino masses at 3$\sigma$ level in sub scheme 1 is confined to the interval between ($0.06 - 3.33 $) eV (illustrated by green region) and the sum of three neutrino masses at 1$\sigma$ level is confined to the interval between ($0.06 - 3.21 $) eV (illustrated by yellow region). 

The bound on mass of sterile neutrino is calculate from Fig. (\ref{fig3}) (b) which is ($0.99 - 4.75$) eV at 3$\sigma$ level and ($1.29-4.65$) eV at 1$\sigma$ level. However, the numerically calculated range for mass of sterile neutrino is more constrained i.e. ($1.04 - 1.55$) eV at 3$\sigma$ level and ($1.31 - 1.32$) eV at 1$\sigma$ level.

Fig. (\ref{fig3}) (c) analysis shows the $m_{\beta\beta}$ for sub scheme 2 (NH) is consistent with present and future experimental data ranges. It can be observed from Fig. (\ref{fig3}) (d) that the sum of four neutrino masses at 3$\sigma$ level is limited to the interval between ($1.15 - 4.81$) eV (depicted by red region) and the sum of four neutrino masses at 1$\sigma$ level is limited to the interval between ($1.42 - 4.69$) eV (depicted by blue region). The sum of three neutrino masses at 3$\sigma$ level in sub scheme 2 (NH) is limited to the interval between ($0.10 - 2.94 $) eV (illustrated by green region) and the sum of three neutrino masses at 1$\sigma$ level is limited to the range between ($ 0.10 - 2.94 $) eV (illustrated by yellow region). The bound on mass of sterile neutrino is calculate from Fig. (\ref{fig3}) (d) which is ($1.03 - 4.71$) eV at 3$\sigma$ level and ($1.30-4.60$) eV at 1$\sigma$ level. However, the numerically calculated range for mass of sterile neutrino is more constrained i.e. ($1.04 - 1.55$) eV at 3$\sigma$ level and ($1.31 - 1.32$) eV at 1$\sigma$ level. Consequently, this theoretical framework demonstrates feasibility in light of both current and future experimental outcomes.

\begin{figure}[H]
\centering
\begin{subfigure}{.5\textwidth}
  \centering
  \includegraphics[width=0.95\linewidth]{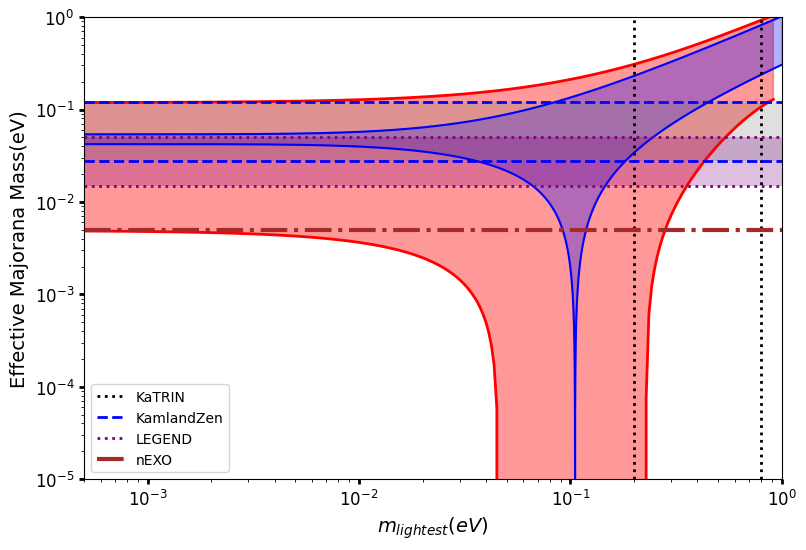}
  \caption{}
 
\end{subfigure}%
\hfill
\begin{subfigure}{.5\textwidth}
  \centering
  \includegraphics[width=1\linewidth]{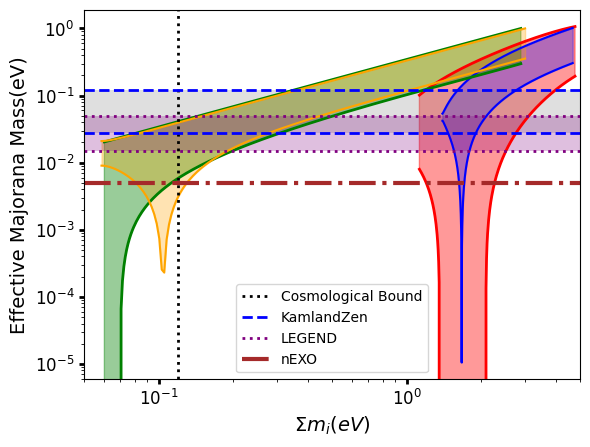}
  \caption{}
  
\end{subfigure}%
\hfill
\begin{subfigure}{.5\textwidth}
  \centering
  \includegraphics[width=1\linewidth]{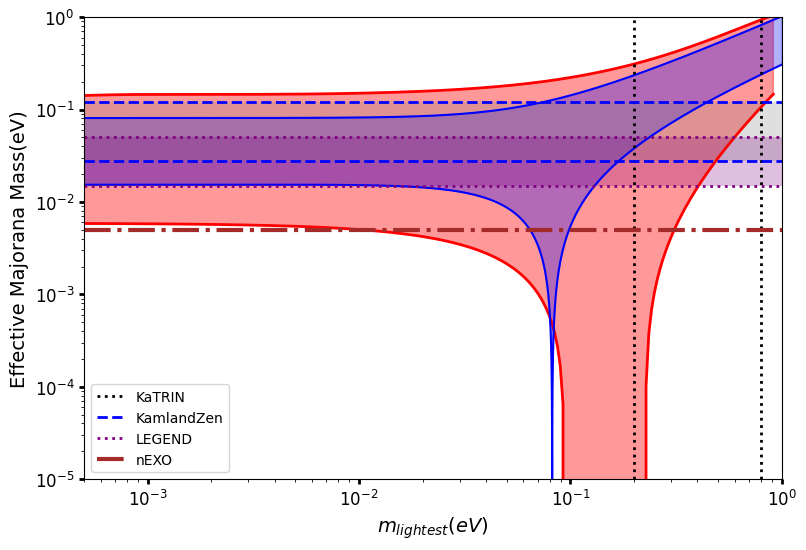}
  \caption{}
  
\end{subfigure}%
\hfill
\begin{subfigure}{.5\textwidth}
  \centering
  \includegraphics[width=1\linewidth]{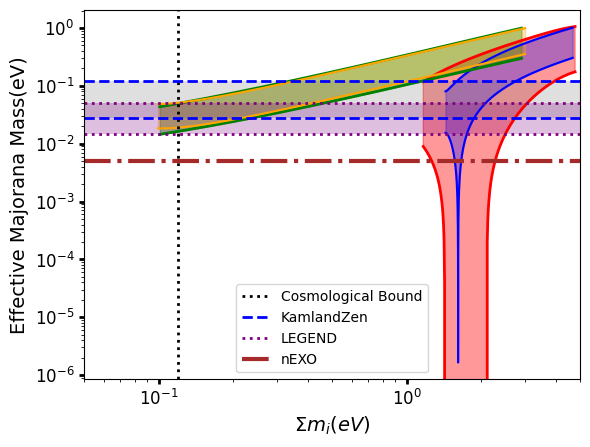}
  \caption{}
 
\end{subfigure}%
\caption{Effective Majorana mass as a function of lightest mass $m_{\text{lightest}}$ (a, c) and sum of masses $\Sigma$ (b, d) for 3+1 scheme (IH) for 1 $\sigma$ and 3 $\sigma$ ranges}
\label{fig6}
\end{figure}

The analysis of Fig. (\ref{fig6}) (a) shows the $m_{\beta\beta}$ for sub scheme 1 (IH) is consistent with present and future experimental data ranges. Similar behavior has been observed in the plots presented in Ref.~\cite{26}.

It can be observed from Fig. (\ref{fig6}) (b) that the sum of four neutrino masses at 3$\sigma$ level is limited to the interval between ($1.14 - 4.78$) eV (depicted by red region) and the sum of four neutrino masses at 1$\sigma$ level is limited to the interval between ($1.35 - 4.62$) eV (depicted by blue region). The sum of three neutrino masses at 3 $\sigma$ level in sub scheme 1 (IH) is confined to the range between ($0.06 - 2.99 $) eV (illustrated by green region) and the sum of three neutrino masses at 1$\sigma$ level is limited to the range between ($ 0.06 - 2.89 $) eV (illustrated by yellow region). The bound on mass of sterile neutrino is calculate from Fig. (\ref{fig6}) (b) which is ($1.02 - 4.72$) eV at 3$\sigma$ level and ($1.23-4.56$) eV at 1$\sigma$ level. However, the numerically calculated range for mass of sterile neutrino is more constrained i.e. ($1.05 - 1.55$) eV at 3$\sigma$ level and ($1.31 - 1.32$) eV at 1$\sigma$ level. 

Fig. (\ref{fig6}) (c) analysis shows the $m_{\beta\beta}$ for sub scheme 2 (IH) for inverted hierarchy is consistent with present and future experimental data ranges. It can be observed from Fig. (\ref{fig6}) (d) that the sum of four neutrino masses at 3$\sigma$ level is limited to the interval between ($1.16 - 4.75$) eV (depicted by red region) and the sum of four neutrino masses at 1$\sigma$ level is limited to the interval between ($1.43 - 4.61$) eV (depicted by blue region). The sum of three neutrino masses at 3$\sigma$ level in sub scheme 2 (IH) is confined to the interval between ($0.09-2.99 $) eV (illustrated by green region) and the sum of three neutrino masses at 1$\sigma$ level is confined to the interval between ($ 0.09 - 2.89 $) eV (illustrated by yellow region). The bound on mass of sterile neutrino is calculate from Fig. (\ref{fig6}) (d) which is ($1.04 - 4.66$) eV at 3$\sigma$ level and ($1.31-4.32$) eV at 1$\sigma$ level. However, the numerically calculated range for mass of sterile neutrino is more constrained i.e. ($1.05 - 1.55$) eV at 3$\sigma$ level and ($1.31 - 1.32$) eV at 1$\sigma$ level. Consequently, this theoretical framework demonstrates feasibility in light of both current and future experimental results. 

We summarize the  limits on $\Sigma (4\nu)$, i.e. sum of four neutrino masses numerically calculated limits on $m_{4}$, limits on three neutrino mass and graphically calculated limit on $m_{4}$ in Table \ref{tab2} and Table \ref{tab3} as

\begin{table}[H]
  \scriptsize 
    \centering
    \begin{tabular}{|c|c|c|c|c|c|c|c|c|c|}
        \hline
        \multirow{2}{*}{Scheme} & \multirow{2}{*}{Ordering} & \multicolumn{4}{c|}{Limits on $\Sigma (4\nu)$ (eV)} & \multicolumn{4}{c|}{Numerically calculated limits on $m_{4}$ (eV)} \\
        \cline{3-10}
         & & \multicolumn{2}{c|}{1 $\sigma$} & \multicolumn{2}{c|}{3 $\sigma$} & \multicolumn{2}{c|}{1 $\sigma$} & \multicolumn{2}{c|}{3 $\sigma$}\\
        \cline{2-10}
         &sub schemes&1&2&1&2&1&2&1&2 \\
        \hline
        3+1 &NH&1.41-4.71&1.42-4.69&1.11-4.81&1.15-4.81&1.31-1.32&1.31-1.32&1.04-1.55&1.04-1.55 \\
        \cline{2-10}
         &IH&1.35-4.62&1.43-4.61&1.14-4.78& 1.16-4.75 &1.31-1.32&1.31-1.32&1.05-1.55&1.05-1.55 \\
        \hline
    \end{tabular}
    \caption{Limits on $\Sigma (4\nu)$ and numerically calculated limits on $m_{4}$ by using third mass squared difference and $m_{\text{lightest}}$ in 3+1 scheme for 1 $\sigma$ and 3 $\sigma$ ranges}
    \label{tab2}
\end{table}

\begin{table}[H]
  \scriptsize  
    \centering
    \begin{tabular}{|c|c|c|c|c|c|c|c|c|c|}
        \hline
        \multirow{2}{*}{Scheme} & \multirow{2}{*}{Ordering} & \multicolumn{4}{c|}{Limits on sum of three neutrino masses (eV)} & \multicolumn{4}{c|}{Graphically calculated limit on $m_{4}$  (eV)} \\
        \cline{3-10}
         & & \multicolumn{2}{c|}{1 $\sigma$} & \multicolumn{2}{c|}{3 $\sigma$} & \multicolumn{2}{c|}{1 $\sigma$} & \multicolumn{2}{c|}{3 $\sigma$}\\
        \cline{2-10}
         & sub schemes  &  1 &  2 & 1 &  2 &  1 &  2 &1 &  2 \\
        \hline
        3+1 & NH & 0.06-3.21 &  0.10-2.94 & 0.06-3.33  & 0.10-2.94 & 1.29-4.65 & 1.30-4.60 & 0.99-4.75  & 1.03-4.71 \\
        \cline{2-10}
         & IH & 0.06-2.89 & 0.09- 2.89 & 0.06-2.99 & 0.09-2.99 & 1.23-4.56&1.31-4.32 & 1.02-4.72 & 1.04-4.66 \\
        \hline
    \end{tabular}
    \caption{Limits on sum of three neutrino masses and graphically calculated limits on $m_{4}$ (difference between the sum of four neutrino masses and three neutrino masses) in 3+1 scheme for 1$\sigma$ and 3$\sigma$ ranges}
    \label{tab3}
\end{table}

\begin{figure}[H]
\centering
\begin{subfigure}{.5\textwidth}
  \centering
  \includegraphics[width=0.95\linewidth]{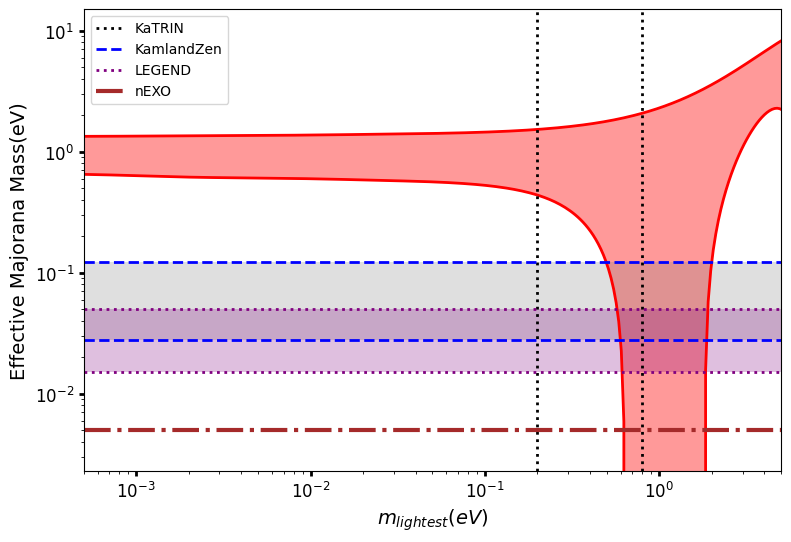}
  \caption{}
 
\end{subfigure}%
\hfill
\begin{subfigure}{.5\textwidth}
  \centering
  \includegraphics[width=1\linewidth]{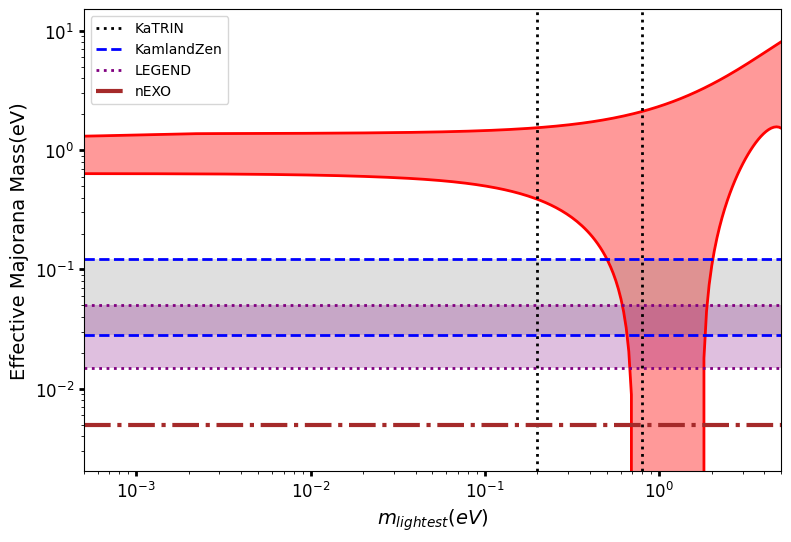}
  \caption{}
 
\end{subfigure}%
\caption{Effective Majorana mass as a function of lightest mass $m_{\text{lightest}}$ for 3+1 scheme (NH) for 3 $\sigma$ ranges with third mass squared difference predicted by the NEUTRINO-4 experiment}
\label{fig7}
\end{figure}

The analysis of $m_{\beta\beta}$ with respect to the lightest neutrino mass is also performed using predictions from the NEUTRINO-4 experiment. The fig.~\ref{fig7}(a) indicates that sub scheme 1 (NH) is consistent with the KamLAND-Zen experimental results. Additionally, the corresponding lightest neutrino mass shows compatibility with the current KATRIN limits, but is not consistent with the proposed sensitivity of the KATRIN experiment. In future, the proposed sensitivity of  KATRIN experiment will rule out this scenario. The sub scheme 2 (NH) in Fig.~\ref{fig7}(b) also exhibits similar results as sub scheme 1.

\begin{figure}[H]
\centering
\begin{subfigure}{.5\textwidth}
  \centering
  \includegraphics[width=0.95\linewidth]{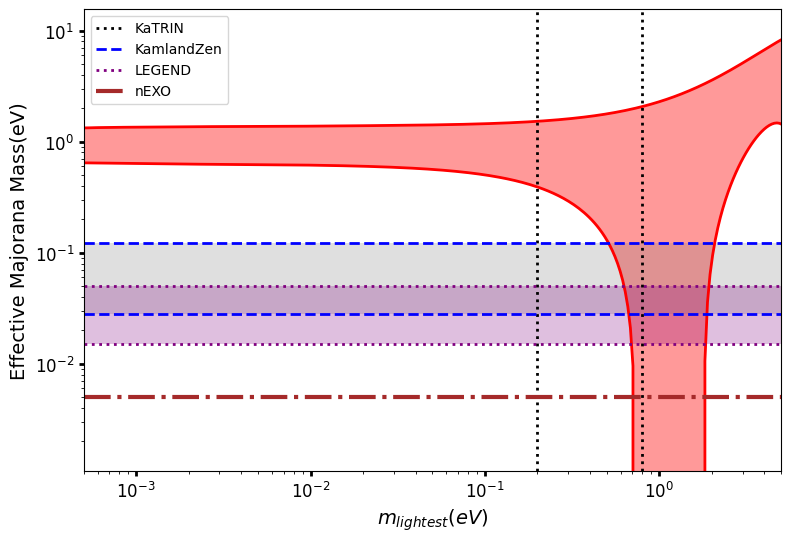}
  \caption{}
 
\end{subfigure}%
\hfill
\begin{subfigure}{.5\textwidth}
  \centering
  \includegraphics[width=1\linewidth]{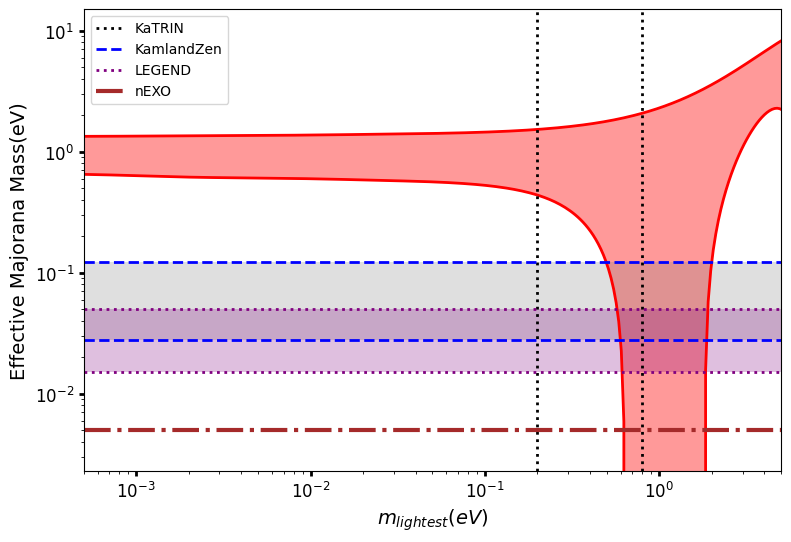}
  \caption{}
 
\end{subfigure}%
\caption{Effective Majorana mass as a function of lightest mass $m_{\text{lightest}}$ for 3+1 scheme (IH) for 3 $\sigma$ ranges with third mass squared difference predicted by the NEUTRINO-4 experiment}
\label{fig8}
\end{figure}

In Fig.~\ref{fig8}, both sub schemes 1 and 2 demonstrate that  \( m_{\beta\beta} \), as a function of the lightest neutrino mass, is compatible with the KamLAND-Zen limits. Moreover, the lightest neutrino mass is compatible with the current KATRIN limits but shows incompatibility with the proposed sensitivity of KATRIN experiment. The NEUTRINO-4 experiment shows insensitivity towards the neutrino mass hierarchy because in both the normal hierarchy and inverted hierarchy the dominant term arises from the third mass squared difference, which is of the same magnitude in both cases. Therefore, the NEUTRINO-4 experiment cannot distinguish between normal and inverted hierarchies of neutrinos.

\section{Conclusions}
In this work, we assumed a sterile neutrino mixed with three active neutrinos and investigated the third mass squared difference in the 3 $\sigma$ range, i.e., $1.1-2.4$ $\text{eV}^2$ and at best fit value, i.e., 1.73 $\text{eV}^2$. 
We used the global fits of the current understanding of neutrino mass squared differences and mixing angles, recent and future experiments on the effective Majorana mass, neutrino absolute mass scale experiments and reactor experiments data in our study.  In the possible different active-sterile schemes such as 3+1, 1+3 and 2+2, we confirmed that recent and future experimental data support 3+1 scheme only in both the normal as well as inverted hierarchies of neutrino masses. As also observed, as in the existing literature, the other two schemes 1+3 and 2+2 are not found favoured by the experimental data. Using the 3+1 scheme, the upper limit on the mass of sterile neutrino is obtained as 4.75 eV for normal hierarchy and 4.72 eV for inverted hierarchy at 3 $\sigma$ level. However, upper limits on the sum of four  neutrino masses are 4.81 eV for NH and 4.78 eV for IH at 3 $\sigma$ level. Since for both the hierarchies the limits of sum of four neutrino masses, thus obtained, are almost similar, therefore, the four neutrino study will not help us in distinguishing between the mass hierarchy of neutrinos. 

In the work, we also investigated the effective Majorana mass in light of the NEUTRINO-4 experiment, which shows that the current experimental bounds provided by KamLAND-Zen are compatible with the effective Majorana mass limits. In this case, the lightest neutrino mass is also compatible with the current limit of the KATRIN experiment. However, at this stage, the proposed sensitivity of the KATRIN experiment rules out this scenario, and this will show significant tension with the stringent bound set by the KamLAND-Zen experiment. As we know, the effective Majorana mass depends on mass eigenvalues, neutrino mixing angles and Majorana phases $\alpha$, $\beta$ and $\gamma$, these parameters are shaping the overall understanding of effective Majorana mass within the proposed schemes. The detailed results and limits on mass of sterile neutrinos are shown in the Table \ref{tab2} and Table \ref{tab3}.

The sterile neutrinos of eV scale are one of the most intriguing possibilities in the search for right handed helicity of neutrinos beyond the Standard Model. Anomalous results from short-baseline experiments have exciting indications, which cannot be explained in the standard framework of $3 \nu$ mixing. 
Furthermore, the 3+1 scheme has been considered as the most promising framework to be tested in the short-baseline neutrino oscillations.

After the emergence the reactor antineutrino anomaly (RAA)\cite{41}, several projects have been proposed to search out the sterile neutrinos of eV mass scale. To resolve the LSND, MiniBooNE and Gallium anomalies, an intense experimental activities have been carried out worldwide. The experiments e.g. NEOS/ Daya Bay\cite{42}, DANSS\cite{43}, STEREO\cite{44}, PROSPECT\cite{45} and SoLiD\cite{47} are already in operation and taking data \cite{48}. The PROSPECT-I aims to measure the signatures of sterile neutrino oscillations, excludes the 3+1 sterile neutrino parameter space above 3~eV$^2$ that was previously unexplored by other neutrino experiments at 95\% C.L \cite{PROSPECT:2024gps}.

The preliminary results are excluding part of the allowed range of parameters, but also giving positive hints for the new oscillation signals. Apart from this, there are several experiments keeping eyes on the signals of 0$\nu\beta\beta$ decay, which have been depicted on the plots shown in the work.  

We are keen to see the exciting results from the new short-baseline neutrino experimental programs and experiments targeting to 0$\nu\beta\beta$ decay in the near future that would enlighten us more on the existence of eV-scale sterile neutrinos and settle the longstanding issue of 0$\nu\beta\beta$ decay.

\section*{Acknowledgements}
The authors are thankful to the Inter University Centre for Astronomy and Astrophysics (IUCAA), Pune for providing necessary facilities during the completion of this work.

\end{document}